\documentclass[12pt]{article}

\usepackage{amsmath}

\begin{document}
\baselineskip 18pt
\begin{center} {\bf \large SOME VARIATIONS\\ ON MAXWELL'S EQUATIONS}
\bigskip
\bigskip
\\ \baselineskip 14pt
Giorgio A. Ascoli \\
Krasnow Institute for Advanced Study,\\
George Mason University,\\
Fairfax, VA 22030, USA\\
{\it ascoli@gmu.edu\/}\\
\bigskip
Gerald A. Goldin \\
Departments
of Mathematics and Physics\\
Rutgers University\\
New Brunswick, NJ 08903, USA\\
{\it gagoldin@dimacs.rutgers.edu\/}\\
{\ }\\
\end{center}

\begin{quote}\baselineskip 14pt
\noindent{\bf Abstract} In the first sections of this article, we
discuss two variations on Maxwell's equations that have been
introduced in earlier work---a class of nonlinear Maxwell theories
with well-defined Galilean limits (and correspondingly generalized
Yang-Mills equations), and a linear modification motivated by the
coupling of the electromagnetic potential with a certain nonlinear
Schr\"odinger equation. In the final section, revisiting an old idea
of Lorentz, we write Maxwell's equations for a theory in which the
electrostatic force of repulsion between like charges differs
fundamentally in magnitude from the electrostatic force of
attraction between unlike charges. We elaborate on Lorentz'
description by means of electric and magnetic field strengths, whose
governing equations separate into two fully relativistic Maxwell
systems---one describing ordinary electromagnetism, and the other
describing a universally attractive or repulsive long-range force.
If such a force cannot be ruled out {\it a priori\/} by known
physical principles, its magnitude should be determined or bounded
experimentally. Were it to exist, interesting possibilities go
beyond Lorentz' early conjecture of a relation to (Newtonian)
gravity.
\end{quote}

\bigskip
\baselineskip 14pt
 \noindent {\it It is a pleasure to dedicate this
paper to G\'erard Emch, whose skeptical perspective helps motivate
those who know him to the pursuit of deeper scientific
understandings.\/}

\newpage
\bigskip
\noindent {\bf \large 1. Introduction\/}

\bigskip \noindent Maxwell's equations are among the most beautiful
in physics, unifying the forces of electricity and magnetism in a
classical field theory that explains electromagnetic waves
\cite{Jackson1999}. Some well-known, profoundly-motivated variations
on Maxwell's equations have included the Born-Infeld theory (a
nonlinear but Lorentz-covariant modification, that introduces an
effective upper bound to the electric field strength), and the
Yang-Mills equations (introducing non-Abelian gauge potentials)
\cite{BornInfeld1934,YangMills1954}. These ideas go back many
decades, and have deeply influenced the development of theoretical
physics. Indeed, there has been a recent resurgence of interest in
non-Abelian Born-Infeld Lagrangians \cite{Roskies1977}, which turn
out to have important application in string theory and related
subjects \cite{Tseytlin1997,Park1999,GaltsovKerner2000,SMK2003}.
More recently, variations of Maxwell's equations have been
considered as ''test theories,'' with respect to which observations
in astrophysics can provide upper bounds to deviations from the
usual equations or laws of physics
\cite{Amelino2005,Laemmerzahl2005}. We nevertheless seek to approach
the idea of modifying Maxwell's equations in new ways with
appropriate humility. None of the variations considered in this
article is {\it ad hoc.\/} Rather, each occurred in answer to a
specific question in fundamental physics.

Sections 2 and 3 review two such modifications considered by the
second author in recent years. The first of these, proposed in joint
work with Vladimir Shtelen
\cite{GoldinShtelen2001,GoldinShtelen2004}, is a class of Galilean
nonlinear Maxwell theories, together with non-Abelian versions that
generalize the Yang-Mills equations and the non-Abelian Born-Infeld
equations. These possibilities arise in answer to the question of
whether and how Maxwell's equations for the four fields $\,{\bf
E}\,$, $\,{\bf B}\,$, $\,{\bf D}\,$ and $\,{\bf H}\,$ can survive
unchanged in the Galilean limit of $\,c \to \infty\,$, a feature
that is present in neither the usual, linear Maxwell theory, nor the
usual Born-Infeld theory. The second variation is a simple, linear
modification, that can be associated with a change over time in some
``constants'' of electromagnetism. This possibility comes up in
answer to the question of how to write gauge-invariant expressions
for the electromagnetic field strengths $\,{\bf E}\,$ and $\,{\bf
B}\,$, when Maxwell's equations are coupled with a natural and very
general family of {\it nonlinear\/} Schr\"odinger time-evolutions in
quantum mechanics \cite{Goldin2000}. Such a family of nonlinear
Schr\"odinger equations was developed and studied in a series of
articles, in joint work with Heinz-Dietrich Doebner and Peter
Nattermann \cite{DoebGol1992, DoebGol1996, DoebGolNatt1999}. In the
general nonlinear equation, a certain ``frictional'' term originally
proposed by Kostin \cite{Kostin1972} is the one requiring a change
in Maxwell's equations for the field strengths, after nonlinear
gauge transformations are taken into account.

Finally, in Section 4, we take up a different question. In a 1900
article that seems to be not very well-known today, Lorentz
explored the idea that (Newtonian) gravity could be explained if
the electrostatic force of repulsion between like charges were
smaller in absolute magnitude than the electrostatic force of
attraction between unlike charges \cite{Lorentz}. Setting aside
Lorentz' conjectured relation to gravity, we want to reopen the
possibility of a difference in magnitude between these forces.
There is then a straightforward and elegant description of the
situation by means of Maxwell's equations, that was partially
written down in Lorentz' original article. Under the given
hypothesis, one may introduce new electric and magnetic fields
whose governing equations separate into two fully relativistic
Maxwell systems---one describing ordinary electromagnetism, and
the other describing an overall attractive or repulsive long-range
force that couples to an ``absolute charge.'' While the latter
force might conceivably have something to do with gravity, it is
more plausible to regard it as an extremely small, but
theoretically possible, correction to ordinary electromagnetism.
Such a correction could, in principle, be time-dependent, and
serve as a further ``test theory'' for astrophysical measurements.
Whether attractive or repulsive, a modified electromagnetism could
be important in modeling the early universe. Unless a known
physical principle rules out such a force {\it a priori,\/} its
magnitude should be regarded as an experimental question.
Well-known nonlinear and non-Abelian generalizations of Maxwell's
equations, and their unification with weak interactions, could
then equally well be constructed from the new equations, opening
up interesting possibilities.

In the remainder of this section we establish notation, summarize
some elementary, familiar background material, and make a few
relevant remarks. All of our discussions pertain to
$(3+1)$-dimensional space-time.

Let us write Maxwell's equations in SI units, as follows
\cite{Jackson1999}:
\begin{equation}
{\bf \nabla} \times {\bf E} =-\frac{\partial \bf B}{\partial
t}\,,\quad {\bf \nabla} \cdot {\bf B} =0\,,\quad {\bf \nabla} \times
{\bf H} =\frac{\partial \bf D}{\partial t} + {\bf j}\,, \quad {\bf
\nabla} \cdot {\bf D} =\rho\,,\label{maxwell}\end{equation}
where $\,{\bf E}({\bf x},t)\,$ is the electric field, $\,{\bf
D}({\bf x},t)\,$ the electric displacement, $\,{\bf B}({\bf x},t)\,$
the magnetic induction, and $\,{\bf H}({\bf x},t)\,$ the magnetic
field; $\,\rho({\bf x},t)\,$ is the charge density, and $\,{\bf
j}\,({\bf x},t)\,$ the electric current density.

The first pair of Eqs. \!(\ref{maxwell}) imply that we can write
$\bf E$ and $\bf B$ in terms of potentials ($\Phi, \ \bf A$),
\begin{equation}
{\bf B}={\bf \nabla} \times {\bf A}\,, \quad {\bf E} =
-\frac{\partial {\bf A}}{\partial t}-{\bf \nabla} \Phi\,.
\label{gifields}
\end{equation}
The choice of $\Phi$ and $ \bf A$ is not unique. For an arbitrary
smooth function $\Theta({\bf x},t)$, new potentials ${\bf A}'$ and
$\Phi\,'$ that are obtained from the {\it gauge transformation\/}
\begin{equation}
{\bf A}'={\bf A} + {\bf \nabla} \Theta\,,\quad
\Phi\,'=\Phi-\frac{\partial \Theta}{{\partial t}} {\label{lingt}}
\end{equation}

\smallskip
\noindent give just the same fields $\,\bf E\,$ and $\,\bf B\,$.
Thus $\,\bf E\,$ and $\,\bf B\,$ are said to be gauge invariant. The
condition $\,\nabla \cdot {\bf B} = 0\,$ expresses the nonexistence
in nature of magnetic monopoles, which in this article we do not
consider changing. Our choice of SI units here is motivated by the
desire to avoid incorporating the speed of light $\,c\,$ into the
definitions of any of the fields, as we shall later be interested in
considering the $\,c \to \infty\,$ limit.

From the second pair of Eqs. \!(\ref{maxwell}), there follows
immediately the equation of continuity,
\begin{equation}
\frac{\partial \rho}{\partial t} + \nabla \cdot {\bf j} \,=\, 0\,,
\label{continuity}
\end{equation}
expressing conservation of electric charge.

Thus far, the system is underdetermined. To complete Maxwell's
equations in the presence of matter, one introduces {\it
constitutive equations\/} relating the fields $\,{\bf D}$, $\,{\bf
H}\,$ to the fields $\,{\bf E}$, $\,{\bf B}$. The usual system of
Maxwell equations {\it in vacuo\/} is obtained using {\it linear\/}
constitutive equations,
\begin{equation}
{\bf D}\,=\,\varepsilon_0{\bf E}\,,\quad {\bf
H}\,=\,\frac{1}{\mu_0}\,{\bf B}\,, \label{linconstitutive}
\end{equation}
where $\,\varepsilon_0\mu_0 = 1/c^2$. However, we shall shortly be
considering a certain class of nonlinear constitutive equations.

One may take the point of view, given the absence of magnetic
monopoles, that the only {\it physically detectable\/} fields are
$\,{\bf E}\,$ and $\,{\bf B}$. These are defined operationally {\it
via\/} the observed Lorentz force $\,{\bf F}\,$ on a small ``test
particle'' with electric charge $\,q\,$ moving with velocity $\,{\bf
v}$:
\begin{equation}
{\bf F} \,=\, q\,{\bf E} + q\,{\bf v} \times {\bf B}\,.
\end{equation}
The fields $\,{\bf H}\,$ and $\,{\bf D}\,$ can then be regarded as
unobservable constructs used to describe, by way of the latter two
Maxwell equations and the constitutive equations, how the observable
fields are {\it produced\/} by charges and currents.

As noted in Ref. \!\cite{GoldinShtelen2001}, we then actually have a
more general class of linear constitutive equations,
\begin{equation}
\left[\begin{array}{cccc}{\bf D}\\
 {\bf H} \end{array}\right] = \left[\begin{array}{cccc}{\varepsilon_0}&\lambda\\
 {-\lambda}&{1/\mu_0} \end{array}\right] \left[\begin{array}{cccc}{\bf E}\\
 {\bf B} \end{array}\right],
 \label{genlinconstitutive}
\end{equation}
which---when combined with Eqs. \!(\ref{maxwell})---lead {\it for
all\/} values of the real parameter $\,\lambda\,$ to the {\it
same\/} set of equations for the observable fields $\,{\bf E}\,$ and
$\,{\bf B}$. Thus the choice $\,\lambda = 0\,$ resulting in Eqs.
\!(\ref{linconstitutive}) is arbitrary.

Also relevant to the forthcoming discussion is the well-known {\it
minimal coupling\/} of the electromagnetic potentials $\,(\Phi, {\bf
A})\,$ with Schr\"odinger's equation for the (complex-valued) wave
function $\,\psi({\bf x},t)\,$ of a single quantum-mechanical
particle having charge \,$q$,
\begin{equation}
    i\hbar\frac{\partial\psi}{\partial t} = \frac{1}{2m}(-i\hbar{{\bf
\nabla}} - q{{\bf A}})^2\psi + q\Phi\psi\,. \label{lse}
\end{equation}
Local $\,U(1)\,$ gauge transformations act on $\,\psi\,$ according
to the formula,
\begin{equation}
\psi^{\,\prime}({\bf x},t) \,=\, e^{\,i \theta ({\bf x},t)}\psi({\bf
x},t)\,,
\end{equation}
and it is easily checked that if $\,\psi\,$ obeys Eq. \!(\ref{lse}),
then $\,\psi^{\,\prime}\,$ obeys a gauge-transformed equation of the
same form, with new electromagnetic potentials given by Eqs.
\!(\ref{lingt}) in which $\,\Theta({\bf x},t) \,=\,
(\hbar/q)\,\theta({\bf x},t)\,$.

The gauge-invariant fields $\,{\bf E}\,$ and ${\bf B}$, which exert
the electric and magnetic forces on the charged quantum particle,
are obtained from $\,\Phi\,$ and $ {\bf A}\,$ using Eqs.
\!(\ref{gifields}), and satisfy Eqs. \!(\ref{maxwell}). The
gauge-invariant probability and probability flux densities for the
particle are given, respectively, by
\begin{equation}
\rho^{\,{\mathrm gi}}=\bar\psi\psi, \quad {\bf J}^{{\mathrm
gi}}=\frac{\hbar}{2im}[\bar\psi{\bf \nabla}\psi-({\bf
\nabla}\bar\psi)\psi]- \frac{q}{m}\,\bar\psi\psi\,{\bf A}\,;
\label{rhoJgilin}
\end{equation}
these also obey an equation of continuity.

Let us remark on the fact that Eqs. \!\!(\ref{maxwell}) respect the
Lorentz transformations of special relativity, while Schr\"odinger's
equation respects Galilean transformations. The minimal coupling of
Eq. \!(\ref{lse}) is compatible with these facts because Eqs.
\!\!(\ref{maxwell}) {\it also\/} respect Galilean transformations.
It is the {\it linear constitutive equations\/} that impose Lorentz
symmetry on the usual Maxwell equations, breaking the Galilean
symmetry (see below).

We reproduce Lorentz transformations here in SI units for
completeness. Let the subscript $\,\parallel\,$ indicate the
component of a vector in the direction of the velocity $\,{\bf v}\,$
of an inertial frame of reference, let the subscript $\,\perp\,$
indicate the component perpendicular to $\,{\bf v}$, and let $v =
|{\bf v}|$. Then with
\begin{equation}
\gamma= \frac{1}{\sqrt{1-v^2/c^2}}\,, \end{equation}
the space-time transformation under the Lorentz boost is
\begin{equation}
{\bf x}\,'_{\parallel}=\gamma({\bf x}_{\parallel} - {\bf v} t)\,,
\quad {\bf x}\,'_{\perp} ={\bf x}_{\perp}\,,
 \quad  t\,'=\gamma(t-\frac{{\bf v} \cdot {\bf x}}{c^2})\,;
 \label{lorentz}
\end{equation}
the field transformations are
\[
{\bf B}\,'_{\parallel}= {\bf B}_{\parallel}\,, \quad  {\bf
B}\,'_{\perp}=\gamma({\bf B}-\frac{1}{c^2}\,{\bf v} \times {\bf
E})_{\perp}\,, \quad {\bf E}\,'_{\parallel}= {\bf E}_{\parallel}\,,
\ {\bf E}\,'_{\perp}=\gamma({\bf E}+{\bf v} \times {\bf
B})_{\perp}\,,
\]
\begin{equation}
{\bf H}\,'_{\parallel}= {\bf H}_{\parallel}\,,\quad {\bf
H}\,'_{\perp}=\gamma({\bf H}-{\bf v} \times \bf D)_{\perp}\,,
\label{fieldtransformations}
\end{equation}
\[{\bf D}\,'_{\parallel}= {\bf D}_{\parallel}, \quad {\bf D}\,'_{\perp}=\gamma({\bf
D}+\frac{1}{c^2}\,{\bf v} \times {\bf H})_{\perp}\,;\]
and the electric current and charge density transformations are
\begin{equation}
{\bf j}\,'_{\,\parallel}=\gamma({\bf j}_{\,\parallel} -\rho\,{v}),
\quad {\bf j}\,'_{\perp} = {\bf j}_{\,\perp}, \quad
\rho\,'=\gamma(\rho -\frac{\bf v \cdot \bf j}{c^2})\,.
\label{currenttransformations}
\end{equation}
The corresponding electromagnetic potential transformations in SI
units are
\begin{equation}
\Phi\,' = \gamma ({\Phi - {\bf v}\cdot {\bf A}})\,, \quad {\bf A}' =
\gamma({\bf A} - \frac{{\bf v}}{c^2}\,\Phi)\,.
\end{equation}
Under Lorentz transformation, the following combinations of the
fields are then invariant:
\[
I_1 ={\bf B}^2 -\frac{1}{c^2}\,{{\bf E}^2}\,, \quad I_2= \bf
B\cdot\bf E\,;\]
\begin{equation}
I_3={\bf D}^2 -\frac{1}{c^2}\,{\bf H^2}\,, \quad  I_4={\bf H} \cdot
\bf D\,; \label{invariants}
\end{equation}
\[ I_5={\bf B} \cdot {\bf H} - {\bf E} \cdot {\bf D}\,, \quad
I_6={\bf B }\cdot {\bf D} + \frac{1}{c^2}\, {\bf E} \cdot {\bf
H}\,.\]
The Born-Infeld Lagrangian as a function of these invariants is
\begin{equation}
 {\mathcal{L}}=1-R, \quad R=
 \frac{b^2}{\mu_0c^2}\sqrt{1+\frac{c^2}{b^2}I_1-\frac{c^2}{b^4}I^2_2}\,\,.
 \label{BI}
 \end{equation}

Next let us write the above in covariant notation. Define
$x^{\mu}=(ct, {\bf x})$, $\mu = 0,1,2,3$, and $x_{\mu}=g_{\mu\nu}
x^{\nu}=(ct, -{\bf x})$, where the metric tensor
$g_{\mu\nu}={\textrm{diag}}\,(1,-1, -1, -1)$, and where summation
over repeated Lorentz indices is understood; for example,
$x_{\mu}x^{\mu}=c^2t^2-{\bf x}^2$. We further define
$\,\partial_{\mu} \equiv
\partial/{\partial x^{\mu}}=[\,(1/c)\,\partial/{\partial t},
 {\bf \nabla}\,]$; and we shall use the antisymmetric Levi-Civita
tensor $\varepsilon^{\,\alpha\beta\mu\nu}$, with
$\varepsilon^{0123}=1$. Then the usual relativistic
 tensor fields $F_{\alpha\beta}$ and ${\cal F}^{\,\alpha\beta}$,
 constructed from the fields
${\bf E}$ and ${\bf B}$, are
\begin{equation}
F_{\alpha\beta}=\left[\begin{array}{cccc}0&(1/c)
 E_1&(1/c)E_2&(1/c)E_3\\-
(1/c)E_1&0&-B_3&B_2\\
-(1/c)E_2&B_3&0&-B_1\\-(1/c)E_3&-B_2&B_1&0
  \end{array}\right],
\label{Fdefs}
\end{equation}
\[
  {\mathcal{F}}^{\alpha\beta}=\frac{1}{2}\,\varepsilon^
 {\alpha\beta\mu\nu}F_{\mu\nu}= \left[\begin{array}{cccc}0&-B_1&-B_2&-B_3
 \\B_1&0&(1/c)E_3&-(1/c)E_2\\ B_2&-(1/c)E_3&0&(1/c)E_1\\
B_3&(1/c)E_2&-(1/c)E_1&0 \end{array}\right],
\]
with
\begin{equation}
F^{\alpha\beta}=g^{\alpha\mu}g^{\beta\nu}F_{\mu\nu}\,,\quad
{\mathcal{F}}_{\alpha\beta}=g_{\alpha\mu}g_{\beta\nu}{\mathcal{F}}^{\mu\nu}\,.
\end{equation}
Likewise,
\begin{equation}
G^{\alpha\beta}=
 \left[\begin{array}{cccc}0&-cD_1&-cD_2&-cD_3\\cD_1&0&-H_3&H_2\\
 cD_2&H_3&0&-H_1\\cD_3&-H_2&H_1&0
 \end{array}\right]\,,  \,\,
 G_{\alpha\beta}=g_{\alpha\mu}g_{\beta\nu}G^{\mu\nu}\,,
\end{equation}
and so forth. Maxwell's equations (\ref{maxwell}) then become
 \begin{equation}
 \partial_{\alpha}{\mathcal{F}}^{\,\alpha\beta}=0; \quad
 \partial_{\alpha}G^{\,\alpha\beta}=j^{\,\beta}, \label{ME}
 \end{equation}
 with $j^{\,\beta}=(c\rho, {\bf j})$. With $A_{\mu}=(\Phi, -{\bf
 A})$, we have from the first
 of Eqs. \!(\ref{ME}),
 \begin{equation}
 {\mathcal{F}}^{\alpha\beta}=\epsilon^{\alpha\beta\mu\nu}
 \partial_{\mu}A_{\nu}\,, \quad F_{\mu\nu}=
 \partial_{\mu}A_{\nu}-\partial_{\nu}A_{\mu}\,. \label{F}
 \end{equation}
 The first two invariants of
 Eqs. \!(\ref{invariants}), that enter Eq. \!(\ref{BI}) for the Born-Infeld
 Lagrangian density, are now written
\begin{equation}
I_1 =\frac{1}{2}F_{\mu\nu} F^{\mu\nu}, \quad I_2 =
-\frac{c}{4}F_{\mu\nu}{\mathcal{F}}^{\mu\nu}.\label{LI}
\end{equation}

Note that our strategy, following Refs.\! \cite{GoldinShtelen2001}
and \cite{GoldinShtelen2004}, has been to postpone writing
constitutive equations for as long as possible. These now relate
$\,{\bf D},\,{\bf H}\,$ to ${\,{\bf E},\,{\bf B}}$; or,
equivalently, they relate $\,G^{\,\alpha\beta}\,$ to
$\,{\mathcal{F}}^{\,\alpha\beta}$. The general form for
Lorentz-invariant constitutive equations is given by \cite{FSS}
 \begin{equation}
 {\bf D} = M{\bf B} + \frac{1}{c^2}N{\bf E}\,, \quad {\bf H}=N{\bf B} - M{\bf
 E}\,,\label{ceq}
 \end{equation}
where $\,M\,$ and $\,N\,$ are functions of the Lorentz invariants in
Eqs. \!(\ref{invariants}), or
\begin{equation}
{\bf B} = R{\bf D} + \frac{1}{c^2}Q{\bf H}, \quad {\bf E}=Q{\bf D} -
R{\bf H}, \label{ceqinverted}
\end{equation}
where $Q$ and $R$ are likewise functions of the invariants. The
linear constitutive equations (\ref{linconstitutive}) correspond to
the choices $M=0, \ N=1/\mu_0$, in which case $\,\varepsilon_0 =
1/\mu_0 c^2$. Other choices lead to nonlinear relativistic field
equations, such as Born-Infeld or Euler-Kockel electrodynamics. It
is natural in Eqs. \!(\ref{ceq}) to take $\,M\,$ and $\,N\,$ to be
functions of just the first two invariants $\,I_1\,$ and $\,I_2\,$
(which depend only on $\,{\bf E}\,$ and $\,{\bf B}\,$), or in the
inverted Eqs. \!(\ref{ceqinverted}) to take $\,R\,$ and $\,Q\,$ to
be functions of just the invariants $\,I_3\,$ and $\,I_4\,$ (which
depend only on $\,{\bf D}\,$ and $\,{\bf H}\,$).

The usual approach to writing nonlinear Maxwell theories is to begin
with the Lagrangian $\,\mathcal{L}\,$, which can be written as a
function of the invariants. The constitutive equations then follow
from the Euler-Lagrange equations. But not all Lorentz-covariant
theories are Lagrangian, and the approach {\it via\/} constitutive
equations is more general.

In covariant form, Eqs. \!\!(\ref{ceq}) become
 \begin{equation}
 G^{\mu\nu}= NF^{\mu\nu}+ cM{\mathcal{F}}^{\mu\nu}
 \equiv M_1\frac{\partial I_1}{\partial F^{\mu\nu}}+ M_2\frac{\partial I_2}{\partial F^{\mu\nu}} ,\label{ce}
 \end{equation}
where $M_1$ and  $ M_2$ are likewise functions of the Lorentz
invariants.

Now we are ready to discuss the three variations to which this
article is devoted.

\bigskip
\bigskip
\noindent {\bf \large 2. Maxwell equations having Galilean limits\/}

\bigskip \noindent
When Maxwell's equations are written for the four fields ${\bf E}$,
${\bf B}$, ${\bf D}$, and ${\bf H}$ as in Eqs. \!(\ref{maxwell}),
the system is underdetermined. In the SI units we are using, these
equations are independent of the speed of light $c$. Furthermore,
the corresponding equations for the primed fields defined by Eqs.
\!(\ref{fieldtransformations}) and the primed currents defined by
Eqs. \!(\ref{currenttransformations}), when written in the primed
coordinates defined by Eqs. \!(\ref{lorentz}), are unchanged from
Eqs. \!(\ref{maxwell})---even though the Lorentz transformations
given in Eqs. \!\!(\ref{lorentz})-(\ref{currenttransformations})
{\it are\/} parameterized explicitly by $c$. Since the invariance of
Eqs. \!(\ref{maxwell}) holds for every value of $c$, we should not
be at all surprised that in the limit $c \to \infty$, these
equations also respect the resulting Galilean transformations,
\[{\bf x}\,' = {\bf x} -{\bf v} t\,, \quad t\,'=t\,,\]
\[{\bf B}\,'= {\bf B}\,, \quad {\bf E}\,'= {\bf E} + {\bf v} \times
{\bf B}\,, \]
\begin{equation}
{\bf H}\,' = \bf H - \bf v \times \bf D\,, \quad {\bf D}\,'= \bf
D\,,
\end{equation}
\[{\bf j}\,' = {\bf j} - \rho \,{\bf v}\,, \quad \rho\,'= \rho\,.\]

Now, the value of $c$ {\it does\/} appear in the constitutive
equations. Thus the choice between Lorentz
 or Galilei symmetry, or the selection of a particular, finite value
 of $c$ under which the Lorentz symmetry holds, resides entirely in the
 constitutive equations.

 For the linear constitutive equations
 (\ref{linconstitutive}) or (\ref{genlinconstitutive}), the speed of light is
 specified by $\,c = (\varepsilon_0 \mu_0)^{-1/2}$. In this case,
 taking a Galilean limit requires that some aspect of Maxwell's
 equations be sacrificed, as discussed in detail by Le Bellac and
 Levy-Leblond \cite{LBLL1973}.

 The constitutive equations
 that select Galilean symmetry, when combined with Eqs. \!(\ref{maxwell}), are
\begin{equation}
{\bf D} = \hat{M} {\bf B}, \quad {\bf H}=\hat{N} {\bf B} -
\hat{M}\bf E,
 \label{giconstitutive}
\end{equation}
 or, equivalently
\begin{equation}
{\bf B} = \hat{R}{\bf D}, \quad {\bf E}=\hat{Q}{\bf D} - \hat{R}{\bf
H}, \label{gicinverted}
\end{equation}
where $\hat{M}$ and  $\hat{N}$, $ \hat{Q}$ and $\hat{R}$ are
arbitrary functions of Galilean invariants,

\[
\hat{I_1} ={\bf B}^2 , \quad \hat{I_2}= \bf B \cdot \bf E;\]
\begin{equation}
\hat{I_3}={\bf D}^2 , \quad  \hat{I_4}=\bf H\cdot\bf D;
\end{equation}
\[\hat{I_5}={\bf B} \cdot {\bf H} - {\bf E} \cdot {\bf D}, \quad
\hat{I_6}=\bf B\cdot\bf D .\]
These  constitutive equations and field invariants are respectively
the formal limits as $c\to\infty$ of their Lorentz invariant
counterparts.

As discussed in Ref. \cite{LBLL1973}, however, taking the
mathematical step of letting $c\to\infty$ is not precisely the same
thing as imposing the low velocity condition $v/c << 1$ on a class
of physical systems governed by the dynamical equations with Lorentz
symmetry. For instance, when ${\bf E}$ and ${\bf B}$ are held fixed,
the limit $c\to\infty$ does not allow the ``electric limit'' of Ref.
\cite{LBLL1973}, although $v/c << 1$ is compatible with it.

Letting $\,{\hat M}\,$ be a constant in Eqs.
\!(\ref{giconstitutive}) requires (since $\nabla \cdot {\bf B} = 0$)
that the charge density $\rho \equiv 0$. Hence the answer to the
question of a consistent Galilean electrodynamics, retaining
Maxwell's equations, the continuity equation, and the Lorentz force,
is a class of  {\it essentially nonlinear} theories, that can arise
as the $c \to \infty$ limit of a class of essentially nonlinear
Lorentz-covariant theories. Indeed, Le Bellac and Levy-LeBlond
emphasize (always assuming linear constitutive equations) the mutual
incompatibility of Galilean invariance, the continuity equation with
non-zero values, and magnetic forces between electric currents.

With nonlinear constitutive equations, these features are no longer
incompatible. Nontrivial choices of $\,{\hat M}\,$ and $\,{\hat
N}\,$ in Eqs. \!\!(\ref{giconstitutive}), or $\,{\hat R}\,$ and
$\,{\hat Q}\,$ in Eqs. \!\!(\ref{gicinverted}), combined with
Maxwell's equations, yield fully consistent Galilean versions of
electrodynamics.

For example, Ref. \!\cite{GoldinShtelen2001} proposes to set
\begin{equation}
\hat{Q} = \frac{1}{\varepsilon}\,\quad \hat{R} \,=\, \alpha \,+\, 2
\alpha^2 \varepsilon\,\frac{{\bf H}\cdot{\bf D}}{|\bf D|^2}\,,
\end{equation}
which are homogeneous functions of the field strengths. This can be
shown to lead to an interesting, albeit non-Lagrangian, theory.

Writing the Lagrangian for a nonlinear relativistic theory as
\begin{equation}
{\mathcal{L}}={\mathcal{L}}(I_1,\, I_2),
\end{equation}
a short calculation in Ref. \!\cite{GoldinShtelen2004} demonstrates
from the Euler-Lagrange equations that
\begin{equation}
N = 2\, \frac{\partial{\mathcal{L}}}{\partial I_1}\,, \quad M =
-\frac{\partial{\mathcal{L}}}{\partial I_2}\,.
\end{equation}
Therefore the necessary compatibility condition for the constitutive
equations to describe such a Lagrangian theory is given by,
 \begin{equation}
 2\, \frac{\partial M}{\partial I_{1}}+ \frac{\partial N}{\partial I_{2}}=0\,.
 \end{equation}
In the Galilean limit, we would take
${\mathcal{L}}={\mathcal{L}}(\hat{I}_1, \,\hat{I}_2)$, and argue
similarly.

The usual Born-Infeld theory does not have a nontrivial Galilean
limit. The Lagrangian ${\mathcal{L}}({I}_1, \,{I}_2)$ is given by
\begin{equation}
 {\mathcal{L}}=1-\mathcal{R}\,, \quad \mathcal{R} =
 \frac{b^2}{\mu_0c^2}\,\sqrt{1+\frac{c^2}{b^2}I_1-\frac{c^2}{b^4}I^2_2}\,,
 \end{equation}
which leads to the constitutive equations (\ref{ceq}) with
\begin{equation}
M(I_1,\,I_2) \,=\, \frac{I_2}{\mu_0 b^2 {\mathcal{R}}}\,, \quad N =
\frac{1}{\mu_0 \mathcal{R}}\,.
\end{equation}
Taking $\,c \to \infty$, we have of course $\,I_1 \to \hat{I}_1 =
\mathbf{B}^2\,$, and $\,I_2 \to \hat{I}_2 =
\mathbf{B}\cdot\mathbf{E}\,$. But for large $\,c$, one has
$\,{\mathcal{R}}\,\approx (c/b) [\hat{I}_1 -
\hat{I}_2^2/b^2]^{1/2}$, whence the limits of $M$ and $N$ are both
zero.

It is therefore suggested in Ref. \!\cite{GoldinShtelen2004} to
modify the Born-Infeld Lagrangian, replacing $\,\mathcal{R}\,$ by
\begin{equation}
\tilde{\mathcal{R}} = \sqrt{1+\frac{c^2}{b^2}\,[\,(1 + \lambda_1
c^2) I_1-\frac{1}{b^2}(1+\lambda_2 c^2) I_2^{\,2}\,]\,}\,,
\label{BImodified}
\end{equation}
where $\lambda_1$, $\lambda_2$ are new constants with the dimensions
of $\,1/c^2$. Now, taking $\,c \to \infty\,$, one obtains the
Galilean constitutive equations (\ref{giconstitutive}), with
\begin{equation}
\hat{M} = \frac{ \lambda_2 \hat{I}_2}{\mu_0 b
\sqrt{\lambda_1\hat{I}_1 - \lambda_2\hat{I}_2^{\,2}/b^2}}\,,\quad
\hat{N} = \frac{b \lambda_1}{\mu_0 \sqrt{\lambda_1\hat{I}_1 -
\lambda_2\hat{I}_2^{\,2}/b^2}}\,. \label{BImodifiedconst}
\end{equation}

Similarly, generalizations of classical (non-Abelian) Yang-Mills
theory are written by means of Lorentz-covariant, nonlinear
constitutive equations. Again, with appropriate choices for the
dynamics, the new systems can have fully consistent
Galilean-covariant limits as $\,c \to \infty$. In analogy with Eqs.
\!(\ref{BImodified})-(\ref{BImodifiedconst}), one obtains a class of
generalizations of non-Abelian Born-Infeld theories that are of this
type  \cite{GoldinShtelen2004}.

\bigskip
\bigskip
\noindent {\bf \large 3. Modification from a nonlinear Schr\"odinger
equation\/}

\bigskip
\noindent Another variation on Maxwell's equations occurs as a
result of considering the coupling of external electromagnetic
fields with nonlinear Schr\"odinger time-evolutions
\cite{Goldin2000,DoebGolNatt1999}. First we write the class of
Schr\"odinger equations under consideration. Refs.
\!\cite{DoebGol1992} and \cite{DoebGol1996} provide extensive
motivation and development, that we omit here; we mainly follow the
discussion in Ref. \!\cite{Goldin2000}.

Letting $\,\psi(\bf{x},t)\,$ be the quantum-mechanical wave
function, and define
\begin{equation}
\hat{\rho}\,({\bf x},t)\,=\,\overline{\psi}\psi\,,\quad \hat{{\bf
j}}\,({\bf x},t) \,=\, (1/2i)\,[\,\overline{\psi} \nabla \psi -
(\nabla \overline{\psi}) \psi\,]\,. \end{equation}
In this article we shall use the notation $\,\hat{\rho} \,({\bf
x},t)\,$ to refer to the spatial probability density for the
quantum-mechanical particle, to distinguish it from the net charge
density $\rho\,({\bf x},t)\,$ that appears in Secs. 1 and 2. Thus
$\,\hat{\rho}\,$ is here the expression we called $\,\rho^{\,\mathrm
gi}\,$ in the first of Eqs. \!(\ref{rhoJgilin}).

Define the real, homogeneous functionals $\,R_1[\psi], \dots,
R_5[\psi]\,$, by
\begin{equation}
R_1 = {\nabla \cdot {\hat{{\bf j}}} \over \hat{\rho}}, \,\,\, R_2 =
\,{\nabla^{\,2} \hat{\rho} \over \hat{\rho}}, \,\,\, R_3 =
{\hat{{{\bf j}}}^{\,2}\over \hat{\rho}^2}, \,\,\, R_4 = {\hat{{\bf
j}} \cdot \nabla\hat{\rho} \over \hat{\rho}^2},\,\,\, R_5 = {(\nabla
\hat{\rho})^2 \over \hat{\rho}^2}. \label{Rjdefs}
\end{equation}
The Laplacian in the linear Schr\"odinger equation (\ref{lse}) can
be expanded with respect to this basis of functionals,
\begin{equation}
\frac{\nabla^2\,\psi}{\psi} \,=\, i R_1[\psi]
\,+\,\frac{1}{2}\,R_2[\psi]\,-\, R_3[\psi] \,-\,
\frac{1}{4}\,R_5[\psi]\,,
\end{equation}
so that it does not appear explicitly in the equation we shall next
write down. The general family of nonlinear Schr\"odinger equations
takes the form,
\[
i\,{{\dot{\psi}} \over {\psi}}\,\,\,\,=\,\,\,\,
i\left[\,\,\sum_{j=1}^2 \nu_j R_j[\psi] \,+\, {{\nabla \cdot ({\cal
A}({\bf x},t)\hat{\rho})}\over \hat{\rho}}\,\right]
\,+\,\,\sum_{j=1}^5 \mu_j R_j[\psi] \,\,\,+\,
\]
\begin{equation}
+\,\,\, U({\bf x},t)\,+\, {{\nabla \cdot ({\cal A}_1({\bf
x},t)\hat{\rho})}\over \hat{\rho}} \,+\, {{{\cal A}_2({\bf x},t)
\cdot \hat{{\bf j}}}\over \hat{\rho}} \,+\, \alpha_1 \,\ln
\hat{\rho} \,+\, \alpha_2\,S\,, \label{nlsegen}
\end{equation}
where $\,\dot{\psi}\,$ is $\,\partial \psi/ \partial t\,$, the
coefficients $\nu_j\,\,(j = 1,2),$ $\mu_j\,\,(j = 1, \dots\,, 5),$
and $\alpha_j\,\,(j = 1,2)$ are all continuously differentiable,
real-valued functions of $\,t\,$; $\,S({\bf x},t)\,$ is
$\,\arg{[\,\psi ({\bf x},t)\,]}\,$; $U({\bf x},t)$ is a real-valued
scalar function; and ${\cal A},\,{\cal A}_1,$ and ${\cal A}_2$ are
distinct real-valued, time-dependent vector fields.

Writing a class of nonlinear Schr\"odinger equation by adding terms
of the form (\ref{Rjdefs}) to the usual, linear Schr\"odinger
equation as in Ref. \!\cite{DoebGol1996}, we have
\begin{equation}
i\hbar \frac{\partial \psi}{\partial t} \,=\, H_{0}\,\psi \,+
\frac{\,i\,}{2} \hbar D\,R_2[\psi]\, \psi \,+\, \hbar \sum_{j=1}^5\,
D^{\,\prime}_j\, R_j[\psi]\,\psi\,, \label{nlse}
\end{equation}
where $\,H_0\,\psi\,$ is given by the right-hand side of Eq.
\!(\ref{lse}), and $\,D\,$ and the $\,D^{\,\prime}_j$ have the
dimension of diffusion coefficients. Then Eq. \!(\ref{nlse}) is
obtained from Eq. \!(\ref{nlsegen}) with the values,
\[
\nu_1 = - \frac{\hbar}{2m}\,,\,\,\, \nu_2 = \frac{1}{2} D\,,\,\,\,
{\cal A} = {q\over 2m} {\bf A}\,,
\]
\[
\mu_1 = D^{\,\prime}_1\,,\,\,\, \mu_2 = - \frac{\hbar}{4m} +
D^{\,\prime}_2\,, \,\,\, \mu_3 = \frac{\hbar}{2m} +
D^{\,\prime}_3\,, \,\,\, \mu_4 =  D^{\,\prime}_4\,, \,\,\, \mu_5 =
\frac{\hbar}{8m} + D^{\,\prime}_5\,,
\]

\smallskip
\noindent
\[
U({\bf x}, t) = \frac{q}{\hbar}\,\Phi\,+\,\frac{q^2}{2m\hbar }\,{\bf
A}^2,\,\,\,\, {\cal A}_1 = 0\,,\,\,\, {\cal A}_2 = -\frac{q}{m} {\bf
A}\,,
\]

\medskip
\noindent
\begin{equation}
\alpha_1 \,=\, \alpha_2 \,=\, 0. \label{numuvals}
\end{equation}

\smallskip
The motivation for the form adopted in writing Eq.
\!(\ref{nlsegen}), for the presence of the terms with
$\,\alpha_1\,$, $\,\alpha_2\,$ and $\,{\cal A}_1\,\neq\,0$, and the
time-dependence of the coefficients, is the possibility of
introducing a group of {\it nonlinear gauge transformations\/} that
leave this family of equations invariant (as a class). With
$\,\psi\,=\,R\exp {[\,iS\,]}$, these take the form $\,\psi\,\mapsto
\,\psi^{\,\prime} \,=\,R^{\,\prime}\exp{[\,iS^{\,\prime}\,]}$, with
\begin{equation}
R^{\,\prime} \,=\, R\,,\quad S^{\,\prime} \,=\, \Lambda S + \gamma
\ln R + \theta\,; \label{RSlaw}
\end{equation}
where $\,\gamma\,$ and $\,\Lambda\,$ are continuously
differentiable, real-valued functions of $\,t\,$, $\,\Lambda
\,\not=\, 0$, and $\,\theta\,$ is a continuously differentiable,
real-valued function of $\,{\bf x}\,$ and $\,t$. Then nonlinear
gauge transformations obey the group law,
\begin{equation}
(\Lambda_1,\,\gamma_1,\,\theta_1)\,
\,(\Lambda_2,\,\gamma_2,\,\theta_2)\,=\, (\Lambda_1\Lambda_2,\,
\gamma_1 \,+\, \Lambda_1\gamma_2,\, \theta_1 \,+\,
\Lambda_1\theta_2)\,. \end{equation}
With quantum-mechanical measurements characterized as sequences of
positional measurements (at distinct times), together with the
application of external fields between positional measurements
\cite{FH1965,Miel1974}, and maintaining the standard interpretation
of $\,\hat{\rho} \,=\, |\psi|^2\,$ as a probability density for the
outcomes of positional measurements, such transformations then leave
the distribution of outcomes of all measurements invariant. They are
also local in space-time, and respect a separation condition for
multiparticle product wave functions \cite{DoebGolNatt1999,GSv1994}.
Notice that with $\,\gamma \equiv 0\,$ and $\,\Lambda\,\equiv 1\,$,
we recover the usual local $\,U(1)\,$ gauge group of
electromagnetism (acting linearly on $\,\psi$) as a subgroup of the
larger group of nonlinear gauge transformations.

Under the nonlinear transformation in Eq. \!(\ref{RSlaw}), we have
\[
\hat{\rho}^{\,\prime} \, =
\,\overline{\psi^{\,\prime}}\,\psi^{\,\prime} \,= \, \hat{\rho}\,,
\]
\begin{equation}
\hat{{\bf j}}^{\,\prime} \, = \,
\frac{1}{2i}\,[\,\overline{\psi^{\,\prime}}\,\nabla \psi^{\,\prime}
- (\nabla \overline{\psi^{\,\prime}}\,)\,\psi^{\,\prime}\,]\, \, =
\, \Lambda\,\hat{{\bf j}}\, + \,\frac{\gamma}{2} \nabla \hat{\rho}
\,+\, \hat{\rho} \nabla \theta\,;\label{rhojtransform}
\end{equation}
so that $\,\hat{\rho}\,$ (as desired) is gauge-invariant (for
nonlinear as well as linear gauge transformations). Thus we shall
also write $\,\hat{\rho} = \rho^{\,\mathrm{gi}}\,$ when we want to
emphasize this. But $\,\hat{{\bf j}}\,$ is not gauge-invariant---one
must write a new gauge-invariant current (see below), to fully
generalize Eqs. \!(\ref{rhoJgilin}). Moreover, if $\,\psi\,$
satisfies Eq. \!(\ref{nlsegen}) then $\psi^{\,\prime}$ likewise
satisfies an equation of this form, but with new (primed)
coefficients and new external fields. The coefficients transformed
under $\,(\gamma,\,\Lambda,\,\theta)\,$ are given by
\medskip
\[
\nu_1^{\,\prime} = \frac{\nu_1}{\Lambda}\,,\quad \nu_2^{\,\prime} =
-\frac{\gamma}{2\Lambda}\nu_1 +\nu_2\,,
\]
\[
\mu_1^{\,\prime} = -\frac{\gamma}{\Lambda}\nu_1 + \mu_1\,,\quad
\mu_2^{\,\prime} = \frac{\gamma^2}{2\Lambda}\nu_1-\gamma \nu_2 -
\frac{\gamma}{2}\mu_1+\Lambda \mu_2\,,
\]

\smallskip
\noindent
\[
\mu_3^{\,\prime} = \frac{\mu_3}{\Lambda}\,,\quad \mu_4^{\,\prime}=
-\frac{\gamma}{\Lambda}\mu_3 + \mu_4\,,\quad \mu_5^{\,\prime} =
\frac{\gamma^2}{4\Lambda}\mu_3 - \frac{\gamma}{2}\mu_4 +
\Lambda\mu_5\,,
\]

\smallskip
\noindent
\begin{equation}
\alpha_1^{\,\prime} \,=\, \Lambda \alpha_1 \,-\, {\gamma \over
2}\,\alpha_2\,+\, {1 \over 2}\,\left({\dot{\Lambda} \over
\Lambda}\gamma \,-\, \dot{\gamma}\right),\quad \alpha_2^{\,\prime}
\,=\, \alpha_2 \,-\,{\dot{\Lambda} \over \Lambda}\,.
\label{primescorr}
\end{equation}
Observe that even if one begins with $\,\alpha_1 = \alpha_2 = 0\,$
and with time-independent coefficients $\,\nu_j\,$ and $\,\mu_j\,$,
the fact that $\,\gamma\,$ and $\,\Lambda\,$ can be time-dependent
{\it requires\/} that in Eq. \!(\ref{nlsegen}) the $\,\alpha_j\,$ be
permitted to take nonzero values, and that all the
$\,\nu_j,\,\mu_j\,$, and $\,\alpha_j\,$ be, in general,
time-dependent. However, the nonlinear gauge transformations do not
mix the coefficients $\,\alpha_j\,$ with the coefficients
$\,\nu_j\,$ and $\,\mu_j\,$.

The nonlinear term with coefficient $\,\alpha_1\,$ was first
proposed as a modification of linear quantum mechanics by
Bialynicki-Birula and Micielski \cite{BM1976}, and the term with
coefficient $\,\alpha_2\,$ was proposed still earlier  by Kostin
\cite{Kostin1972}. The term with coefficient $\,\nu_2\,\neq 0\,$ was
derived by Doebner and Goldin from considerations of local current
algebra representations \cite{DoebGol1992}, which led to the
generalized equation containing the nonlinear functionals $\,R_j\,$
\cite{DoebGol1996}.

The external fields of Eq. \!(\ref{nlsegen}) transformed under
$\,(\gamma,\,\Lambda,\,\theta)\,$ are given in Ref.
\cite{Goldin2000} by
\medskip
\noindent
\[
{\cal A}^{\,\prime} \,=\, {\cal A} \,-\,
\frac{\nu_1}{\Lambda}\,\nabla \theta\,,
\]

\smallskip
\noindent
\[
{\cal A}_1^{\,\prime} \,=\, \Lambda \, {\cal A}_1\, -\, \gamma \,
{\cal A} \,-\, {\gamma \over 2}\,{\cal A}_2 \,+\, \left(\,{\gamma
\over \Lambda}\,\nu_1 - \mu_1 + {\gamma \over \Lambda}\, \mu_3 -
\mu_4\right)\,\nabla \theta\,,
\]

\medskip
\noindent
\[
 {\cal A}_2^{\,\prime} \,=\,
{\cal A}_2 - \frac{2\mu_3}{\Lambda}\,\nabla \theta\,,
\]
\[
U^{\,\prime} \,=\,\Lambda\,U \,-\, \dot{\theta}
\,+\,\left({\dot{\Lambda} \over \Lambda} - \alpha_2\right)\theta
\,+\,{{\mu_3} \over \Lambda}\,[\,\nabla \theta\,]^{\,2} \,+\,
\quad\quad\quad\quad\quad\quad
\]
\smallskip
\noindent
\begin{equation}
\quad\quad\quad\quad \left(\mu_4 - \mu_3\,{\gamma \over
\Lambda}\,\right) \,\nabla^2\theta \,+\,{\gamma \over
2}\,\nabla\cdot{\cal A}_2 \,-\,{\cal A}_2 \cdot \nabla \theta.
\label{potprimes}
\end{equation}

\bigskip \noindent
Observe that Eqs. \!(\ref{primescorr})-(\ref{potprimes}) imply both
the nonzero $\,{\cal A}_1\,$ and nontrivial $\,{\cal A}_2\,$ values
are required in Eq. \!(\ref{nlsegen}). Even if one begins with
$\,{\cal A}_1 \,\equiv\, 0\,$ and $\,{\cal A}_2 \,\equiv\, -2\,{\cal
A}\,$, as in the linear Schr\"odinger equation [{\it cf.\/} Eqs.
\!(\ref{numuvals})], the nonlinear gauge transformations compel one
to introduce more general values for these fields. Nonlinear
Schr\"odinger equations with arbitrary values of $\,{\cal A}_2\,$
were proposed by Haag and Bannier \cite{HB1978}, while the
interaction with a general external vector field $\,{\cal A}_1\,$
was considered in Ref. \cite{G1997}.

Next let us write the equations of motion described by this class of
nonlinear Schr\"odinger equations entirely in terms of
gauge-invariant quantities, as in Ref. \cite{Goldin2000}---where
``gauge invariance'' is interpreted with respect to the group of
{\it nonlinear\/} gauge transformations. To start, a gauge-invariant
current density $\,{\bf J}^{\mathrm{gi}}\,$ may be written
\begin{equation}
\,{\bf J}^{\mathrm{gi}}\,=\, -\,2\nu_1\,\hat{\\{\bf j}} \,-\,2 \nu_2
\nabla \hat{\rho} \,-\, 2 \hat{\rho} {\cal A}\,, \label{Jgi}
\end{equation}
which evidently reduces to Eq. \!(\ref{rhoJgilin}) for the linear
Schr\"odinger equation [\,when $\,\nu_1 = - \hbar/2m,\,\,\nu_2 =
0,\,\, {\cal A} = (q/2m){\bf A}\,$]. We thus have (again) a
continuity equation for the probability density, $\,\partial
\rho^{\,\mathrm{gi}}/\partial t \,=\, - \nabla \cdot {\bf
J}^{\mathrm{gi}}$. Refs. \cite{DoebGol1996} and
\cite{DoebGolNatt1999} provide a set of gauge-invariant parameters
necessary for the desired description,
\[
\tau_1 = \nu_2 - \frac{1}{2}\mu_1\,,\quad \tau_2 = \nu_1\mu_2
-\nu_2\mu_1\,,\quad \tau_3 = \frac{\mu_3}{\nu_1}\,,\quad \tau_4 =
\mu_4 - \mu_1 \frac{\mu_3}{\nu_1}\,,
\]
\[
\tau_5 = \nu_1\mu_5 - \nu_2\mu_4 +
\nu_2^{\,2}\,\frac{\mu_3}{\nu_1}\,,\quad
\]
\begin{equation}
\beta_1 \,=\, \nu_1 \, \alpha_1 \,-\, \nu_2 \,\alpha_2 \,+\, \nu_2\,
{\dot{\nu}_1 \over \nu_1} - \dot{\nu}_2\,,\quad \,\beta_2 \,=\,
\alpha_2 \,-\,  {\dot{\nu}_1 \over \nu_1}\,. \label{gaugeinvariants}
\end{equation}
When $\tau_1 \not= 0$, $\tau_4 \not= 0$, or $\beta_2 \not= 0$,
time-reversal invariance is violated. When $\tau_3 \not= -1$ or
$\tau_4 \not= 0$, Galilean invariance is violated.

Gauge-invariant external magnetic and electric fields are now given
by the formulas \cite{Goldin2000}
\smallskip
\noindent
\[
{\cal B} = \nabla \times {\cal A}\,= \,{q \over 2m}\,{\bf B},
\]
\begin{equation}
{\cal E} \,=\, - \,\nabla\hat{U} - \frac{\partial {\cal A}}{\partial
t}\,-\, \beta_2\,{\cal A}\,=\, {q \over 2m}\,{\bf E}\,, \label{Enew}
\end{equation}
where
\begin{equation}
\hat{U} \,=\,-\,\nu_1\,U \,-\, \tau_3\,{\cal A}^{\,2} \,-\, (\tau_4
- 2\tau_1\tau_3)\,\nabla \cdot {\cal A} \,+\,{\cal A}\cdot{\cal A}_2
\,-\, \nu_2\,\nabla \cdot {\cal A}_2\,. \label{Uhatcorrected}
\end{equation}
Thus $\,\hat{U}\,$ is to be identified with $\,(q/2m)\Phi\,$ (which
may be directly checked for the linear Schr\"odinger equation); but
the main point here is that the formula for $\,{\bf E}\,$ in terms
of $\,\Phi\,$ and $\,{\bf A}\,$ has been modified from Eq.
\!(\ref{gifields}) to include an extra term, originating with
Kostin's nonlinearity:
\begin{equation}
\,{\bf E}\,=\,-\nabla \Phi \,-\, \frac{\partial {\bf A}}{\partial t}
\,-\, \beta_2\,{\bf A}\,. \label{Edefnew}
\end{equation}
The extra term is {\it necessary\/}---if we leave it out of Eq.
\!(\ref{Enew}), $\,\mathcal{E}\,$ fails to be gauge-invariant. We
also have new gauge-invariant external vector fields,
\smallskip \noindent
\begin{equation}
{\cal A}_1^{\,\,\mathrm{gi}} =\nu_1 {\cal A}_1 +
\left(\frac{2\nu_2\mu_3}{\nu_1} - \mu_1 - \mu_4\right){\cal A} -
\nu_2{\cal A}_2\,, \,\,\,\,\, {\cal A}_2^{\,\,\mathrm{gi}} = {\nu_1
\over 2\mu_3}\,{\cal A}_2 - {\cal A}\,. \label{gaugeinvA}
\end{equation}

Continuing to follow Ref. \cite{Goldin2000}, we are now in a
position to write equations of motion obtained from Eq.
\!(\ref{nlsegen}), in a form that is manifestly gauge-invariant for
the group of nonlinear gauge transformations. We use the
hydrodynamical variables variables $\,\rho^{\,\mathrm{gi}}$ and
$\,{\bf J}^{\mathrm gi}/\rho^{\,\mathrm{gi}}\,$, where the latter
has the interpretation of a gauge-invariant velocity field. We then
have the interpretation of the (gauge-invariant) magnetic induction
as a vorticity field,
\begin{equation}
\nabla \times \,\left(\,\frac{{\bf J}^{\mathrm
{gi}}}{\rho^{\,\mathrm{gi}}}\,\right)\,=\,-2\,{\cal B}\,=\,{q \over
m}\,{\bf B}\,.
\end{equation}
\label{curlJgioverrhogi}
The dynamical equations are the equation of continuity that we wrote
above,
\begin{equation}\frac{\partial \rho^{\,\mathrm{gi}}}{\partial t}\,=\, - \nabla
\cdot {\bf J}^{\mathrm{gi}}\,,\end{equation}
together with the equation for the changing velocity field,
\[
{\partial \over {\partial t}} \left(\,{{\bf J}^{\mathrm{gi}} \over
\rho^{\,\mathrm{gi}}}\,\right) \,=\, \nabla \left[ \,2\tau_1\,\nabla
\cdot \left(\,{{{\bf J}^{\mathrm{gi}}} \over \rho^{\,\mathrm{gi}}}\,
\right) \,+\, 2 \tau_2\,{{\nabla^{\,2} \rho^{\,\mathrm{gi}}} \over
\rho^{\,\mathrm{gi}}} \,+\,{1 \over 2}\,\tau_3\left(\,{{{\bf
J}^{\mathrm{gi}}} \over \rho^{\,\mathrm{gi}}}\, \right)^2\,\right]
\]
\[
\,+\,\nabla \left[ \,(\,2\tau_1\,[1 + \tau_3]\,- \tau_4\,)
\left(\,{{{\bf J}^{\mathrm{gi}}} \over \rho^{\,\mathrm{gi}}}\,
\right)\,\cdot \,{{\nabla \rho^{\,\mathrm{gi}}} \over
\rho^{\,\mathrm{gi}}}\, \,+\,2\tau_5 \left({\nabla
\rho^{\,\mathrm{gi}}} \over
{\rho^{\,\mathrm{gi}}}\right)^2\,\,\right]
\]

\[
\,+\,\,\nabla \left[\,\,2\, {\nabla \cdot {({\cal
A}_1^{\,\,\mathrm{gi}}\rho^{\,\mathrm{gi}})} \over
\rho^{\,\mathrm{gi}}} \,-\,2\tau_3\,{\cal
A}_2^{\,\,\mathrm{gi}}\cdot \left(\,{{{\bf J}^{\mathrm{gi}}} \over
\rho^{\,\mathrm{gi}}}\, \right) \,+\,2 \beta_1 \,\ln
\rho^{\,\mathrm{gi}}\,\,\right]
\]

\begin{equation}
\,-\, \beta_2\,\left(\,{{{\bf J}^{\mathrm{gi}}} \over
\rho^{\,\mathrm{gi}}}\, \right) \,+\, {q \over m}\,{\bf E}\,.
\label{gimotion}
\end{equation}

\smallskip \noindent In Eq. \!(\ref{gimotion}) we see that
$\,\beta_2\,$, taken to be positive, has a natural interpretation as
a gauge-invariant coefficient of friction---it governs the magnitude
of the term in $\,\partial_t\,({\bf
J}^{\mathrm{gi}}/\rho^{\,\mathrm{gi}})$ that is proportional to
${\bf J}^{\mathrm{gi}}/\rho^{\,\mathrm{gi}}$.

In the framework of the nonlinear quantum mechanics discussed here,
the (gauge-invariant) expected values for the position, velocity,
and acceleration of the quantum particle (all of which are functions
of $t$) are given respectively by the following expressions:
\[
<{\bf x}> \,=\, \int {\bf x}\,\rho^{\,\mathrm{gi}}({\bf x})\,d{\bf
x}\,,
\]

\begin{equation}
<{\bf v}> \,=\, {{\partial <{\bf x}>} \over {\partial t}} \,=\, \int
\rho^{\,\mathrm{gi}}\,\left(\,{{{\bf J}^{\mathrm{gi}}} \over
\rho^{\,\mathrm{gi}}}\, \right)\,d{\bf x}\,=\, \int {\bf
J}^{\mathrm{gi}}({\bf x})\,d{\bf x}\,, \label{forcelaws}
\end{equation}

\[
<{\bf a}> \,\,=\,\, {{\partial <{\bf v}>} \over {\partial t}}
\,\,=\,\,\]
\[
\int \rho^{\,\mathrm{gi}} \,\left[ \,{1 \over 2}\,\nabla
\left(\,{{{\bf J}^{\mathrm{gi}}} \over \rho^{\,\mathrm{gi}}}\,
\right)^2 \,+\,\left(\,{{{\bf J}^{\mathrm{gi}}} \over
\rho^{\,\mathrm{gi}}}\, \right)\,\times \,{q \over {m}}\,{\bf B}\,
\,+\,{\partial \over {\partial t}} \left(\,{{{\bf J}^{\mathrm{gi}}}
\over \rho^{\,\mathrm{gi}}}\, \right)\,\,\right]\,d{\bf x}\,.
\]

\smallskip
\noindent In Eqs. \!(\ref{gimotion})-(\ref{forcelaws}), we see that
the laws of force describing the interaction of the charged particle
with the $\,{\bf E}\,$ and $\,{\bf B}\,$ fields are unchanged from
those in linear quantum mechanics.

Now Eq. \!(\ref{Edefnew}) gives us the variation on Maxwell's
equations that is the focus of this section. The usual equations for
$\,{\bf E}\,$ and $\,{\bf B}\,$ are replaced by
\begin{equation}
\,\nabla \times {{\bf E}}\,=\, -\,{\partial\,{{\bf B}} \over
\partial t} \,-\, \beta_2 \,{\bf B}\,, \quad  \nabla \cdot {\bf B} =
0\,. \label{Maxnew}
\end{equation}
Note that the second of these equations is still consistent with the
first. Let us take $\,\beta_2\,$ to be a constant, independent of
$\,t\,$. If $\,{\bf E_0}({\bf x},t)\,$, $\,{\bf B_0}({\bf x},t)\,$
satisfy the original Maxwell equations (with $\beta_2 = 0$), then
fields satisfying Eqs. \!(\ref{Maxnew}) are given by
\begin{equation}
{\bf E} = {\bf E_0}\,e^{\,-\beta_2 t},\quad {\bf B} = {\bf
B_0}\,e^{\,-\beta_2 t}. \label{Maxnewsolns}
\end{equation}
But nonlinear quantum mechanics alone does not specify the remaining
two Maxwell equations. One possibility is to maintain the
constitutive equations \!(\ref{linconstitutive}), with fixed
coefficients $\,\varepsilon_0\,$ and $\,\mu_0\,$. Then taking
$\,{\bf E_0}\,$, $\,{\bf B_0}\,$, $\,{\bf D}_0 = \varepsilon_0 {\bf
E}_0\,$, $\,{\bf H}_0\,=\,(1/\mu_0){\bf B}_0$, $\,\rho_{\,0}\,$ and
$\,{\bf j}_{\,0}\,$ to satisfy the usual, linear Maxwell equations
\!(\ref{maxwell}), we must have
\begin{equation}
\quad {\bf D} = {\bf D_0}\,e^{\,-\beta_2 t},\quad {\bf H} = {\bf
H_0}\,e^{\,-\beta_2 t}\,,
\end{equation}
and
\[
\nabla \cdot {\bf D} \,=\, \rho\,,\,\,\mathrm{with} \quad \rho \,=\,
\rho_{\,0} \,e^{\,-\beta_2 t}\,,
\]
\begin{equation}
\nabla \times {\bf H} \,=\, \frac{\partial \bf D}{\partial t} \,+\,
\beta_2 \,{\bf D} \,+\, {\bf j}\,, \,\,\mathrm{with} \quad {\bf j}
\,=\, {\bf j}_{\,0} \,e^{\,-\beta_2 t}\,.
\end{equation}
That is, with $\,\beta_2\,> 0\,$, the magnitudes of all the electric
charges and currents are decaying exponentially with time. Of
course, the equation of continuity for $\,\rho\,$ and $\,{\bf j}\,$
no longer holds, and net charge is no longer conserved. Instead, we
have the equation
\begin{equation}
\frac{\partial \rho}{\partial t} \,=\, - \nabla \cdot {\bf j} -
\beta_2 \rho\,.
\end{equation}
The nonzero value for $\,\beta_2\,$, interpreted as a coefficient of
friction experienced by a charged particle, has introduced a
preferred universal reference frame. The corresponding Maxwell
theory is no longer covariant.

Another alternative is to join the {\it standard\/} Maxwell
equations for $\,{\bf D}\,$ and $\,{\bf H}\,$ with Eqs.
\!(\ref{Maxnew}) by supposing that the coefficients in the
constitutive equations---the permittivity and permeability of free
space---are time-dependent. With
\begin{equation}
\varepsilon(t) \,=\,\varepsilon_0\,e^{\,+\beta_2 t}\,, \quad \mu(t)
\,=\,\mu_0\,e^{\,-\beta_2 t}\,,
\end{equation}
we may combine Eqs. \!(\ref{Maxnew}) and \!(\ref{Maxnewsolns}) with
the constitutive equations,
\begin{equation}
{\bf D}\,=\,\varepsilon (t){\bf E}\,,\quad {\bf H}\,=\,\frac{1}{\mu
(t)}\,{\bf B}\,. \label{tdconstitutive}
\end{equation}
Then $\,{\bf D} \,=\, {\bf D}_0\,$, $\,{\bf H} \,=\, {\bf H}_0\,$,
$\,\rho \,=\, \rho_0\,$, and $\,{\bf j} \,=\, {\bf j}_{\,0}\,$. We
have $\,\varepsilon(t)\mu(t) = 1/c^2\,$ for all $\,t\,$. The
equation of continuity holds for $\,\rho\,$ and $\,{\bf j}\,$, and
the net charge is conserved.

Since current systems of units {\it define\/} the vacuum
permittivity to have a fixed numerical value, the modification
embodied in Eqs. \!\!(\ref{tdconstitutive}) requires some
reexamination of the way in which we define our units of measurement
for electromagnetism. We shall return briefly to the idea of
time-dependent permittivity and permeability toward the end of the
next section.

\bigskip
\bigskip
\noindent {\bf \large 4. Force differences between like and unlike
charges\/}

\bigskip
\noindent For our final variation on Maxwell's equations, we pursue
the suggestion by Lorentz that the electrostatic force of repulsion
between like charges is slightly different in magnitude from the
electrostatic force of attraction between unlike charges. While
Lorentz proposed this idea in 1900 as a way to arrive at a
universally attractive (Newtonian) gravitational force, we want to
regard it here as just a modification of classical electrodynamics.
Let us write a description of the situation in terms of field
strengths.

The electrostatic forces among pairs of positively-charged point
particles having charges $\,Q_+ > 0\,$ and $\,q_+ > 0\,$ and/or
negatively-charged point particles $\,Q_- < 0\,$ and $\,q_- < 0\,$,
is described by
\[
{\bf F}_\ell \,=\, \frac{1}{4\pi\varepsilon_\ell}\,\frac{Q_+
q_+}{r^2}\,{\bf \hat{r}} \,=\,
\frac{1}{4\pi\varepsilon_\ell}\,\frac{Q_- q_-}{r^2}\,{\bf
\hat{r}}\,,
\]
\begin{equation}
{\bf F}_u \,=\, \frac{1}{4\pi\varepsilon_u}\,\frac{Q_-
q_+}{r^2}\,{\bf \hat{r}} \quad \mathrm{or}\quad {\bf F}_u \,=\,
\frac{1}{4\pi\varepsilon_u}\,\frac{Q_+ q_-}{r^2}\,{\bf \hat{r}}\,,
\label{electrostatic}
\end{equation}
where $\,\hat{{\bf r}}\,$ is the unit vector at the location of each
particle, pointing away from the other, and $\,r\,$ is the distance
between the pair. The subscripts $\,\ell\,$ and $\,u\,$ stand for
``like'' and ``unlike'' respectively, and
$\,\varepsilon_\ell\,\neq\,\varepsilon_u\,$. If $\,|Q_+| = |Q_-|\,$
and $\,|q_+| = |q_-|\,$, then the forces $\,{\bf F}_u \,$ in the
second equation are equal.

Introduce the electric field $\,{\bf E}_+\,$ exerting force on
positive charges, and the electric field $\,{\bf E}_-\,$ exerting
force on negative charges; also the displacement fields $\,{\bf
D}_+\,$ and $\,{\bf D}_-\,$ produced (respectively) by positive and
negative charges. In ordinary electromagnetism, $\,{\bf E} = {\bf
E}_+ = {\bf E}_-\,$, while $\,{\bf D} =\,{\bf D}_+ + {\bf D}_-\,$.
Evidently we must keep track separately of the density
$\,\rho_+\,\geq 0\,$ of positive charge and the density
$\,\rho_-\,\leq 0\,$ of negative charge. From Eqs.
\!(\ref{electrostatic}), the force $\,{\bf F}_+\,$ experienced by
the positively charged particle with charge $\,q_+\,$ in the
presence of a composite having charges $\,Q_+\,$ and $\,Q_-\,$ at a
distance $r$ is given by
\begin{equation}
{\bf F}_+\,=\,q_+ {\bf E}_+ \,=
\,q_+\left(\frac{1}{4\pi\varepsilon_\ell}\,\frac{Q_+}{r^2}\,+
\,\frac{1}{4\pi\varepsilon_u}\,\frac{Q_-}{r^2}\,\right){\bf
\hat{r}}\,,
\end{equation}
and similarly for a negatively charge particle,
\begin{equation}
{\bf F}_-\,=\,q_- {\bf E}_- \,=
\,q_-\left(\frac{1}{4\pi\varepsilon_u}\,\frac{Q_+}{r^2}\,+
\,\frac{1}{4\pi\varepsilon_\ell}\,\frac{Q_-}{r^2}\,\right){\bf
\hat{r}}\,.
\end{equation}
There is a net electrostatic force between neutral composites that
is attractive if $\,\varepsilon_u\,<\,\varepsilon_\ell\,$, and
repulsive if $\,\varepsilon_u\,>\,\varepsilon_\ell\,$. The situation
thus far is described by the Maxwell equation
\begin{equation}
\nabla \cdot \left[\begin{array}{cccc}{\bf D}_+\\
 {\bf D}_- \end{array}\right] = \left[\begin{array}{cccc}\rho_+\\
 \rho_- \end{array}\right],
 \label{divDpm}
\end{equation}
and the constitutive equation
\begin{equation}
\left[\begin{array}{cccc}{\bf E}_+\\
 {\bf E}_- \end{array}\right] = \left[\begin{array}{cccc}{{1}/\varepsilon_\ell}&{1/\varepsilon_u}\\
 {1/\varepsilon_u}&{1/\varepsilon_\ell} \end{array}\right] \left[\begin{array}{cccc}{\bf D}_+\\
 {\bf D}_- \end{array}\right].
 \label{constitutiveEpm}
\end{equation}

Similarly, introduce $\,{\bf B}_+\,$ and $\,{\bf B}_-\,$ as the
magnetic inductions that exert velocity-dependent forces on positive
and negative moving charges $q_+$ and $q_-$ (respectively); so that
the total forces are given (respectively) by
\begin{equation}
F_+ \,=\,q_+{\bf E}_+ \,+\,q_+ {\bf v} \times {\bf B}_+\,,\quad F_-
\,=\,q_-{\bf E}_- \,+\,q_- {\bf v} \times {\bf B}_-\,.
\label{newforcelaws}
\end{equation}
Evidently, we must also keep track separately of the current of
positive charge $\,{\bf j}_{\,+}$ and the current of negative charge
$\,{\bf j}_{\,-}$. We introduce $\,{\bf H}_+\,$ and $\,{\bf H}_-\,$
as the magnetic fields {\it produced\/} (respectively) by electric
currents $\,{\bf j}_{\,+}$ and $\,{\bf j}_{\,-}$, and (respectively)
by changing displacement fields $\,{\bf D}_+\,$ and $\,{\bf D}_-\,$.
Then we obtain, consistent with Lorentz covariance and the absence
of magnetic monopoles, the additional Maxwell equations,
\[
\nabla \times \left[\begin{array}{cccc}{\bf H}_+\\
 {\bf H}_- \end{array}\right] = \frac{\partial}{\partial t}\left[\begin{array}{cccc}{\bf D}_+\\
 {\bf D}_- \end{array}\right] + \left[\begin{array}{cccc}{\bf j}_{\,+}\\
 {\bf j}_{\,-} \end{array}\right],
\]
\begin{equation}
\nabla \times \left[\begin{array}{cccc}{\bf E}_+\\
 {\bf E}_- \end{array}\right] = -\frac{\partial}{\partial t}\left[\begin{array}{cccc}{\bf B}_+\\
 {\bf B}_- \end{array}\right],\quad \nabla \cdot \left[\begin{array}{cccc}{\bf B}_+\\
 {\bf B}_- \end{array}\right] = 0\,,
 \label{maxwellpm}
\end{equation}
with the constitutive equation
\begin{equation}
\left[\begin{array}{cccc}{\bf B}_+\\
 {\bf B}_- \end{array}\right] = \left[\begin{array}{cccc}{\mu_\ell}&{\mu_u}\\
 {\mu_u}&{\mu_\ell} \end{array}\right] \left[\begin{array}{cccc}{\bf H}_+\\
 {\bf H}_- \end{array}\right],
 \label{constitutiveBpm}
\end{equation}
where
\begin{equation}
\varepsilon_\ell \mu_\ell \,=\, \varepsilon_u \mu_u \,=\,
\frac{1}{c^2}\,. \label{epslumuluc2}
\end{equation}
From Eq. \!(\ref{divDpm}) and the first of Eqs. \!(\ref{maxwellpm}),
we have separate continuity equations for $\,\rho_+\,$, $\,{\bf
j}_{\,+}\,$ and $\,\rho_-\,$, $\,{\bf j}_{\,-}\,$,
\begin{equation}
\frac{\partial}{\partial t}\left[\begin{array}{cccc}\rho_+\\
 \rho_- \end{array}\right]\,+\, \nabla \cdot \left[\begin{array}{cccc}{\bf j}_{\,+}\\
 {\bf j}_{\,-} \end{array}\right] \,=\, 0\,.
 \label{continuitypm}
\end{equation}

Thus far we have written in Eqs. \!(\ref{divDpm}) and
(\ref{maxwellpm}) a {\it doubled\/} set of Maxwell equations,
indexed by $\,(+,-)\,$, that are coupled by the matrix constitutive
equations (\ref{constitutiveEpm}) and (\ref{constitutiveBpm}). To
recover the theory as a perturbation of ordinary classical
electrodynamics, {\it define\/} $\,\rho = \rho_+ + \rho_-\,$ and
$\,{\bf D} = {\bf D}_+ + {\bf D}_-\,$; so that $\,\rho\,$ is the
{\it net\/} charge density, and $\,\nabla \cdot {\bf D} = \rho$.
Defining $\,{\bf E} = \frac{1}{2}({\bf E}_+ + {\bf E}_-)$, we
recover the constitutive equation $\,{\bf E} = (1/\varepsilon_0){\bf
D}\,$ by setting
\begin{equation}
\frac{1}{\varepsilon_0} =
\frac{1}{2}\left(\frac{1}{\varepsilon_\ell}+\frac{1}{\varepsilon_u}\right).
\label{eps0def}
\end{equation}
Similarly, defining $\,{\bf j} = {\bf j}_{\,+} + {\bf j}_{\,-}\,$
and $\,{\bf H} = {\bf H}_+ + {\bf H}_-\,$, we have the usual Maxwell
equation $\,\nabla \times {\bf H} = \partial {\bf D}/\partial t
\,+\, {\bf j}\,$. Letting $\,{\bf B} = \frac{1}{2}({\bf B}_+ + {\bf
B}_-)$, we also have the Maxwell equations $\,\nabla \times {\bf E}
= - \partial {\bf B}/\partial t\,$ and $\,\nabla \cdot {\bf B} = 0$,
and we recover the constitutive equation $\,{\bf B} = \mu_0{\bf
H}\,$ by setting
\begin{equation}
\mu_0 = \frac{1}{2}\left(\mu_\ell + \mu_u\right). \label{mu0def}
\end{equation}
Then it follows from Eqs. \!(\ref{epslumuluc2}), (\ref{eps0def}),
and (\ref{mu0def}) that $\,\varepsilon_0 \mu_0 = 1/c^2\,$. In short,
we still have the four fields $\,{\bf E}\,$, $\,{\bf B}\,$, $\,{\bf
D}\,$, and $\,{\bf H}\,$, and they still satisfy the usual Maxwell
equations incorporating the net charge density and net electric
current density, with coefficients $\,\varepsilon_0\,$ and $\,\mu_0
= 1/\varepsilon_0 c^2\,$.

But we also have an {\it additional\/} set of fields, density, and
current,
\[
{\bf \widetilde{D}} \,=\, {\bf D}_+ - {\bf D}_-\,,\quad {\bf
\widetilde{E}} \,=\, \frac{1}{2}\left({\bf E}_+ - {\bf E}_-\right),
\]
\begin{equation}
{\bf \widetilde{H}} \,=\, {\bf H}_+ - {\bf H}_-\,,\quad {\bf
\widetilde{B}} \,=\, \frac{1}{2}\left({\bf B}_+ - {\bf B}_-\right),
\end{equation}
\smallskip
\[
{\widetilde{\rho}} \,=\, {\rho}_+ - {\rho}_-\,,\quad {\bf
\widetilde{j}} \,=\, {\bf j}_+ - {\bf j}_-\,.
\]

\medskip \noindent
These {\it also\/} obey Maxwell's equations; but with new constants
in their constitutive equations, $\,\widetilde{\varepsilon}\,$ and
$\,\widetilde{\mu}\,$, that are given by
\begin{equation}
\frac{1}{\widetilde{\varepsilon}} =
\frac{1}{2}\left(\frac{1}{\varepsilon_\ell}-\frac{1}{\varepsilon_u}\right),\quad
\widetilde{\mu} = \frac{1}{2}\left(\mu_\ell - \mu_u\right).
\end{equation}
In ordinary electromagnetism, $\,\widetilde{\varepsilon}\,$ is
infinite and $\,\widetilde{\mu}\,$ is zero---so that, although
$\,{\widetilde{\rho}}\,$, $\,{\bf \widetilde{j}}\,$, $\,{\bf
\widetilde{D}}\,$ and $\,{\bf \widetilde{H}}\,$ are defined and
nontrivial, $\,{\bf \widetilde{E}} \equiv 0\,$ and $\,{\bf
\widetilde{B}} \equiv 0\,$.

If it is not actually infinite, the magnitude of
$\,\widetilde{\varepsilon}\,$ is presumably very large compared with
that of $\,\varepsilon_0\,$; while if it is not zero, the magnitude
of $\,\widetilde{\mu}\,$ is small compared with that of $\,\mu_0\,$.
One verifies straightforwardly that
$\,\widetilde{\epsilon}\,\widetilde{\mu} = 1/c^2\,$; but the new
constants may be of either sign---both positive (for a net
attractive force between neutral composites), or both negative (for
a net repulsive force). We thus have obtained a pair of fully
decoupled Maxwell systems, consistent with the equations set down in
Lorentz' original article.

An idealized composite point particle, having positive charge $\,q_+
\geq 0\,$ and negative charge $\,q_- \leq 0\,$, may be equivalently
described as having {\it net\/} charge $\,q = q_+ + q_-\,$ and {\it
absolute\/} charge $\,\widetilde{q} = q_+ - q_- \geq 0\,$. Such a
particle, moving with velocity $\,{\bf v}\,$, experiences according
to Eqs. \!(\ref{newforcelaws}) the total force
\begin{equation}
{\bf F} \,=\, {\bf F}_+ + {\bf F}_-\,=\,q\left({\bf E} \,+ \,{\bf v}
\times {\bf B}\right)\,+\,\widetilde{q}\,({\bf \widetilde{E}} \,+
\,{\bf v} \times {\bf \widetilde{B}})\,.
\end{equation}
Thus the new fields $\,\widetilde{{\bf E}}\,$, $\,\widetilde{{\bf
B}}\,$, $\,\widetilde{{\bf D}}\,$, and $\,\widetilde{{\bf H}}\,$
couple to the {\it absolute charge\/} (which is always positive),
and the {\it absolute current;\/} while the usual fields $\,{\bf
E}\,$, $\,{\bf B}\,$, $\,{\bf D}\,$, and $\,{\bf H}\,$ still exist
in this framework and couple to the {\it net charge\/} and the
{\it net current.\/} One should no longer automatically take
positive charge flowing to the right to be indistinguishable
mathematically or physically from the same amount of negative
charge flowing to the left. The net currents ${\bf j}$ are the
same in these two cases, but the absolute currents
$\widetilde{{\bf j}}$ are equal and opposite.

Let us close this section with some comments and speculations, many
of them rather obvious, about such a ``doubled electromagnetism''
theory.

In his original article \cite{Lorentz}, Lorentz took the new force
to be attractive. He sought to identify the absolute charge with
mass (and consequently the absolute current with momentum), and to
calculate whether the precession of the perihelion of Mercury's
orbit could then be understood as due to the (very small) magnetic
force that would originate from the absolute current. He concluded
that the resulting force would be too weak to explain the
astronomical observations, and of course this line of thinking was
superseded by the success of Einstein's general relativity.

However, we want to entertain the idea of a modified
electrodynamics that does not identify the extra fields with
gravity, and that treats absolute charge not as mass but as an
additional property of matter. Of course, this does not preclude
the possibility that existing measurements of gravitational forces
have erroneously incorporated a small extra electrostatic force
(attractive or repulsive). It would seem to be an especially
interesting conjecture that the new force is repulsive. In any
case, we appear to have an additional parameter with which to fit
cosmological models, and an additional ``test theory'' for study
through observations in astrophysics.

Since we have a new set of fields obeying Maxwell's equations, we
would also need to have a new type of electromagnetic wave
(coupling weakly with the absolute charge), a new type of photon,
and a new quantum electrodynamics (see below).

Now the absolute electric charge of a system must be at least
equal to the net electric charge, and at least equal to the sum of
the absolute charges of the system's components. However, it could
in principle be greater. The absolute charge of a nucleon, for
example, might be the sum of the absolute values of the charges of
its constituent quarks; but one could also conjecture additional,
unobserved positive and negative charges in equal measure,
contributing to a larger value of the overall absolute charge.
While net charge is quantized in fixed units, it is plausible but
not necessary that absolute charge be similarly quantized. Thus,
it does not appear to be inconsistent to take the absolute charge
to be proportional to the mass, as Lorentz implicitly did.

However, modern particle physics offers no fundamental theoretical
reason to make such an assumption. If we make reference only to
constituent quarks and leptons, the absolute charge of a proton
(comprised of two up quarks and one down quark) is $5/3$; that of
a neutron (comprised of one up quark and two down quarks) is
$4/3$; and that of an electron (taken to be fundamental) is $1$.
Then the absolute charge of a proton together with an electron is
$8/3$, double that of the neutron, while the respective masses are
very close to equal. Under these assumptions, the absolute charge
per gram of electrically neutral matter comprised of heavier
elements is macroscopically different from that of matter
comprised of lighter elements, and their accelerations under the
Earth's absolute electric field would be different in magnitude.
Since such differences are not observed, we should take the
empirically-determined magnitude of any new inverse-square-law
force of ``absolute electromagnetism'' to be small compared with
Newtonian gravity---making it {\it extremely\/} small in
comparison with ordinary electromagnetism.

Still, one may conjecture that physical ``constants'' are not
actually constant, but change as the universe ages. We discussed
in Sec. \!3 the possibility of the vacuum permittivity and
permeability changing exponentially with $\,t\,$, producing a
modification in Maxwell's equations and a universal frictional
force that breaks covariance. One may instead take these to be
fixed at $\,\varepsilon_0\,$ and $\,\mu_0\,$, but entertain the
possibility that $\,\varepsilon_u\,$ and $\,\varepsilon_\ell\,$
are changing, and that they have not always been as close as they
are today. For example, we could have
\begin{equation}
\tilde{\varepsilon}(t) = \tilde{\varepsilon}_0\,e^{\,+\beta
t}\,,\quad \tilde{\mu}(t) = \tilde{\mu}_0\,e^{\,-\beta t}\,,
\end{equation}
and modify the Maxwell equations for the perturbing fields
$\,\widetilde{{\bf E}}\,$, $\,\widetilde{{\bf B}}\,$,
$\,\widetilde{{\bf D}}\,$, and $\,\widetilde{{\bf H}}\,$ as in
Sec. 3.

Furthermore, should there be regions of space-time containing
plasmas of electrons and positrons, these might contribute
proportionally more to absolute charge than to gravitational mass.
Such speculations leave open some possibilities for observable
effects in astrophysics, even for a small force.

While net charge and absolute charge are both conserved when there
is no particle creation or annihilation, it seems clear that
absolute charge is {\it not\/} conserved by fundamental particle
processes. [Of course, at the time of Lorentz' paper, mass,
positive charge, and negative charge would all have been taken as
separately conserved.] As long as we stay with Lorentz' idea that
absolute charge is proportional to mass, then (as mass is
transformed into energy during fundamental particle processes) we
have a conservation law. But if we take absolute charge to be an
independent quantity with which the new electromagnetic fields
couple, it becomes an unlikely, speculative possibility that
absolute charge transforms into something previously unknown
during annihilation processes, so as to maintain a conservation
law. More likely, one should take the new $\,U(1)\,$ gauge
symmetry to be broken outside the classical domain that is
governed by Maxwell's equations, requiring a different quantum
electrodynamics for absolute electromagnetism.

We have not discussed the question of absolute charge from virtual
particle-antiparticle pairs, or vacuum polarization. It appears
that, unlike the situation for net charge, it should be possible
to have a coherent superposition of quantum states having
different absolute charges. For example, the neutral pion, written
as a linear combination of up and down quark-antiquark pairs,
$\,(\bar{u}u\,-\,\bar{d}d)/\sqrt{2}\,$, would combine states of
absolute charges $\,4/3\,$ and $\,2/3\,$ (in units of the
electron's charge).

If Lorentz' conjecture breaks no known physical principle, then
the question of a discrepancy in magnitude between the
electrostatic forces between like and unlike charges is purely an
empirical one, and the best we can do {\it in principle\/} is to
establish an experimental upper bound to this discrepancy (or,
equivalently, to the ratio
$\,\varepsilon_0/\widetilde{\varepsilon}\,$). For example, it is
clear that his conjecture continues to respect the charge
conjugation invariance of electromagnetism. However, it should be
noted that current physics tends to {\it assume\/} the equality in
magnitude between these forces. Thus the present, official
definition of the coulomb is effectively as a unit of {\it net\/}
charge, while the value of $\,\varepsilon_0\,$ is not measured but
fixed by definition. Just as we have come to distinguish
(theoretically) gravitational mass from inertial mass, and to
regard their proportionality as a question to be determined by
experiment, so may it be necessary to distinguish (theoretically)
$\,\varepsilon_\ell\,$ from $\,\varepsilon_u\,$, and to regard
their closeness as having a value to be bounded by experiment.

Possible further development of a ``doubled electromagnetism''
theory could entail its involvement in electroweak unification or
in the standard model, its nonlinear modifications (as discussed
for Maxwell's equations in Sec. \!\!2 of this article), its
coupling with linear and nonlinear Schr\"odinger quantum mechanics
(as discussed for Maxwell fields in Sec. \!3 of this article), its
non-Abelian generalizations, and its quantum electrodynamics.

\bigskip
\bigskip
\noindent {\bf \large  Acknowledgments\/}

\bigskip
\noindent One of the authors (G. Goldin) thanks H.-D. Doebner,
J.~Lucido, V.~Shtelen, and G.~Svetlichny for interesting
discussions on topics included in this article.

\end{document}